\documentstyle[prl,aps,epsf]{revtex}
\begin{document}
\twocolumn[\hsize\textwidth\columnwidth\hsize\csname @twocolumnfalse\endcsname

\title{Polymer Reptation in Disordered Media }
\author{Dinko Cule and Terence Hwa}
\address{Department of Physics, University of California 
at San Diego, La Jolla, CA 92093-0319}
\date{\today }
\maketitle

\begin{abstract}

The effect of ambient disorders and sequence
heterogeneities on the reptation of a long polymer is studied with the
aid of a disordered tube model.
The dynamics of a random heteropolymer is found to be much slower than
that of a homopolymer due to collective pinning effects.
The asymptotic properties belong to the universality class 
of a directed path
in (1+1)-dimensional random media.

\vspace{10pt}
\vspace{10pt}
\end{abstract}
] 
The stochastic motion of a polymer chain entangled in a disordered medium
such as a gel is of much scientific and technological interest. A convenient
starting point for understanding the polymer dynamics is 
de Gennes' reptation model, which describes the polymer's wormlike
motion along a fictitious tube threaded through an array of fixed obstacles~%
\cite{reptation,dg0,de}. In the original reptation model, the polymer was
assumed to be homogeneous, and effects due to randomness in the medium
(e.g., spatial variations in the pore size of a gel) were neglected.
A later argument by Harris~\cite{harris} showed that the static
configuration of a
self-avoiding polymer is in fact not affected by the randomness, a result
supported by numerical simulations~\cite{lee,muthu0}. However, extensive
numerical~\cite{muthu0,zimm,slater} and 
experimental~\cite{hoag,paa} studies found the {\em dynamics} of polymers 
in random media to be much slower than the classical reptation 
dynamics~\cite{reptation,dg0,de}. These results are understood
qualitatively within an ``entropic trapping" framework~\cite{muthu0,machta}.

It should be noted that although
some of the best known applications of reptation theory are concerned with
the behavior of biopolymers which are inherently {\em heterogeneous}, there has
been little theoretical work on the reptation of heteropolymers beyond a
restricted model analyzed by de Gennes many years ago~\cite{dg}. 
In this article, we address the combined effect of the heterogeneity 
of the polymer and the randomness of the media.
Polymer motion is separated into two components: reptation within a tube
and diffusive motion of the center-of-mass. For a homopolymer,
the disorder does not affect the reptational motion, but the c-o-m motion
is drastically slowed down due to entropic trapping. For a heteropolymer, 
the reptational motion is also drastically affected: The 
well-known algebraic reptation time becomes exponentially long, making the
dynamics of the heteropolymer even much slower.
Our results are obtained with the aid of a disordered tube model, which we
find to be  analogous to the well-known problem of a directed path in
random media~\cite{dp}.

Consider first  a long, flexible
homopolymer entangled in a covalently-crosslinked gel matrix, characterized
by a typical pore size $a$. We shall take
$a > \xi \gg b$, $\xi $ being the persistence length of the polymer 
and $b$ being the monomer size.
(E.g., double-stranded DNA in agarose gel has $a \sim
5000\AA$, $\xi \approx 500\AA$ and $b = 3.4 \AA$~\cite{tinland}; pore sizes
are typically much smaller in polyacrylamide gels~\cite{paa}.) 
Since the equilibrium configuration of the polymer
is that of a self-avoiding walk, described by the size exponent
$\nu \gtrsim 1/2$ even in the presence of disorders~\cite{harris,lee,muthu0},
we shall describe the dynamics of the polymer
by the reptation tube model~\cite{de}. In this model, the polymer 
is confined to a tube  of diameter $\sim a$~\cite{reptation},
and modeled as  ``beads" linearly connected by (entropic)
springs of spring constant $K=\frac 32k_BT/\xi ^2$.
Configurational entropy associated with the exponential number of 
(self-avoiding) tube trajectories results in an effective tensile force
acting on the two ends of the bead-spring chain as in the
disordered-free  case~\cite{de}. This effect is incorporated
into the model by letting the springs acquire a finite
equilibrium length of the order $a$~\cite{de}. Each link of this bead-spring
chain therefore represents an elementary ``blob'' of 
$M_{\rm blob}\sim a^2/(b\xi)$ monomers whose physical size is $\sim O(a)$.
 Confinement of the polymer
to the tube costs a free energy (per blob) of the order $\overline{V}=k_BT\cdot
(\xi /a)^2$~\cite{dg0}. Since the diameter 
of the reptation tube embedded in the
disordered gel matrix is non-uniform along its length, we describe the
variable confinement entropy by a ``random potential'' $V(s)$, where $s$
denotes the curvilinear coordinate along the tube. For
a typical self-avoiding tube trajectory, 
$V$ may be modeled by a short-range correlated
random variable, characterized by the variance $\overline{\delta V(s)\delta
V(s^{\prime })}=\Delta _V\delta _a(s-s^{\prime })$, where $\delta V(s)
\equiv V(s)-\overline{V}$, $\delta _a(s)$ 
is an exponentially damped function of range
$a$, and the overline denotes average over the ensemble of tubes. Assuming
that the fractional variance in tube diameter variation is $O(1)$, we have
$\Delta _V\sim \overline{V}^2$.

A key advantage of the original tube model~\cite{de} is that the nonlocal
excluded-volume interaction between the beads can be neglected.
This results from the entropy-generated stretching forces, which makes
the chain {\em extended} in the tube, i.e., the  contour length of the chain
being linearly proportional to the number of beads.
Consequently, the large scale dynamics of the polymer is obtained
simply as the 1d Rouse dynamics of the beads.
Let us assume that the nonlocal interaction between the beads (confined
in the self-avoiding tube) can be neglected also for the disordered case.
Then the statistical mechanics
of a long polymer with degree of polymerization $M \gg M_{\rm blob}$ 
is  given by a bead-spring chain of $N \simeq M/M_{\rm blob} \gg 1$ links, 
with the ``Hamiltonian'' 
\begin{equation}
{\cal H}_0=\sum_{n=0}^N\left\{ \frac K2(s_{n+1}-s_n-a)^2+V(s_n)\right\} ,
\label{homo}
\end{equation}
where $s_n$ denotes the coordinate of the $n^{th}$ bead along the tube.

The legitimacy of neglecting the nonlocal bead-bead interaction certainly
deserves scrutiny: As in the disorder-free case, the approximation is 
justified provided the chain is in an extended conformation. On the other hand,
a Gaussian chain in a tube can in principle collapse~\cite{muthu1}
to a region where $\delta V$ is large and negative, corresponding to a
section of the tube with wider openings. If the chain described by 
(\ref{homo}) collapses, then the tube  model  would not be self-consistent.
What opposes the collapse of the Gaussian chain is again
the tensile stretching forces generated by the configurational entropy
of the self-avoiding tube.
This physics is reflected in the model (\ref{homo}) as a competition between
the ``elastic energy" cost against chain collapse, of the order
$\frac{K}{2}a^2$ per spring, and the disorder energy gain, of the 
order $\sqrt{\Delta_V}$. The outcome of this competition has been obtained
recently in the context of ``non-Hermitian quantum mechanics"~\cite{nelson}: 
A discontinuous phase transition between the collapsed and
the extended state is found at a certain critical point given by
the condition $\frac K2a^2\sim \sqrt{\Delta _V}$. 
Using the expressions given above for $K$ and $\Delta _V$ for the polymer
problem at hand, we find that {\em the extended state is always preferred}
under the presumed condition $a\gtrsim \xi $.
Thus, the disordered tube model (\ref{homo}) is justified self-consistently. 
It will be used from here on to generate the large scale dynamics of
self-avoiding polymers in random media.

Classical reptation dynamics is characterized by the reptation time
$\tau_{R,0}\sim N^3$, which is the time it takes 
for the polymer to reptate from
a given tube to a completely different tube in the adjacent neighborhood. 
For time scales much exceeding $\tau_{R,0}$, the polymer behaves as
a point particle undergoing Brownian motion. 
This leads to a diffusion coefficient $D_0(N)\sim N^{2\nu}/\tau_{R,0}(N)$. 
In the presence of randomness, the reptation time
$\tau_R$ is modified (see below); 
but more significantly, the large scale motion becomes
one of thermally-activated barrier hopping, since adjacent tubes may
be narrower and hence higher in energy~\cite{machta}. 
The typical barrier height $E_b(N)$
can be taken as the free energy variation of a polymer confined to
cells of size $\sim N^\nu$. This variation is bounded 
to be within the order of $\pm N^{1/2}$ and
is  found numerically~\cite{machta1} to scale as $N^\omega$,
with $\omega\approx 0.15$ in 3d and $\omega \approx 0.28$ in 2d.
Thus the overall diffusion coefficient is reduced drastically to 
$D \sim [N^{2\nu}/\tau_R(N)] e^{-E_b(N)/k_B T}$ for large $N$.
 
To estimate  the reptation time $\tau_R(N)$ itself for the
random system, it is necessary to separate the 
sliding motion of a polymer along its prescribed tube
from the thermally-activated barrier hopping. This can be accomplished
within the disordered tube model (\ref{homo}) by
artificially imposing a {\em periodic boundary condition} on the
random potential $V$, i.e., $V(s)=V(s+N\,a)$. With this simplification,
the extended state of the chain   can be
 obtained straightforwardly,
by observing a discrete translational symmetry $s_n\rightarrow
s_{n+1}$ which the model (\ref{homo}) possesses even in the presence of
disorder: The polymer can move along the tube by the propagation of a
longitudinal ``defect'' or ``kink'' (e.g., a configuration with $s_n=s_{n+1}$%
) much like what was proposed in the original reptation model~\cite{reptation}.
Thus, the random potential
is irrelevant in the extended state~\cite{nelson},
and the scaling
properties in the sliding regime are the same as those in the disorder-free
case. For example, $\tau_R \sim \tau_{R,0}$, accompanied by
a contour length fluctuation $\delta L_0(N)\approx ((k_BT/K)N)^{1/2}$. 

The above analysis clarifies the role of entropic trapping for a homopolymer:
The very slow motion of a long homopolymer in
random media is dominated by the exponentially long waiting time 
$\sim \tau_{R,0} \, e^{N^\omega/k_BT}$
needed to overcome large spatial variations in the 
{\em mean} confinement potential. 
It is not a result of local ``bottlenecks''
which impedes the reptational motion along the tube.
To study this behavior experimentally or numerically, it is necessary
to have sufficiently long polymers such that each polymer threads
through {\em many} local bottlenecks.

We next consider a heteropolymer which may carry side
groups of vastly different sizes, or may have different degrees of partial
charges, hydrophobicities, etc. 
Following de~Gennes~\cite{dg}, we
model the coarse-grained effect of the side groups by assigning a ``charge'' 
$q_n\in \{\pm Q\}$ to each bead $n$.
Then the interaction energy 
$V$ in Eq.~(\ref{homo}) becomes explicitly $n$-dependent, with the form 
\begin{equation}
V(s_n,n)=V_0(s_n)+q_nV_1(s_n),  \label{U}
\end{equation}
where $V_0(s)\pm QV_1(s)$ is respectively the free energy cost of placing a
bead of $\pm $ charge at the coordinate $s$ along the tube. We will take the
random potentials $V_0$ and $V_1$ to be short-range correlated, 
with variances of the order $\Delta _V$. To model a
heteropolymer, we choose the charges $\left\{ q_n\right\} $ randomly, with
probability $p$ for $q=+Q$ and $1-p$ for $q=-Q$, such that $[q]=(2p-1)\,Q$,
and $[\delta q_m\,\delta q_n]=\Delta _q\,\delta _{m,n}$, $\Delta
_q=4p(1-p)\,Q^2$. (Here, $[...]$ denotes average over the ensemble of random
sequences.) The magnitude of $Q$ depends on the
coarse-graining scale $a$ and variations of the monomer-tube
interaction. Let the latter be characterized 
by a fractional variance $\sigma$,
then we have $Q^2= \sigma M_{\rm blob}$.

As in the homopolymer case, a Gaussian chain described by the
Hamiltonian (\ref{homo}) with the interaction (\ref{U}) may be in either the
collapsed or the extended state depending on the magnitude of the elastic
energy $\frac K2a^2$. However for a RHP, different monomers tend to prefer
different sections of the tube. Hence the
extended state is even more preferred, and we are again justified
to neglect the non-local bead-bead interaction.

Let us consider now the reptational motion of an extended RHP, again 
by imposing periodic boundary condition on the $V$'s. Heterogeneity
in sequence composition has profound effects on the dynamics,
as the random $q_n$'s remove the translational symmetry $s_n\to s_{n+1}$
described above for the homopolymer. Thus, larger beads
prefer to occupy segments of the tube with wider cavities, etc.
De~Gennes demonstrated the existence
of such a heteropolymeric effect by considering a singular limit of the RHP
model~\cite{dg}, where
the longitudinal elasticity of the polymer is suppressed by taking
$K\to \infty $. Below, we will show
that a significant heteropolymer effect can exist generically 
for an elastic RHP.

To proceed, we introduce a longitudinal displacement field, $u_n=s_n-n\,a
$, and write the Hamiltonian as 
\begin{equation}
{\cal H}=\sum_{n=1}^N\left\{ \frac K2(u_{n+1}-u_n)^2+W(u_n,n)\right\} ,
\label{dp1}
\end{equation}
where $W(u_n,n)=\delta q_n\cdot \delta V_1(u_n+na,n)$. [There are actually
two additional terms in ${\cal H}$: $\sum_nq_n\overline{V_1}$ and $\Sigma
_n\left( V_0(s_n)+[q]V_1(s_n)\right) $. The first term is simply  an
overall energy shift and does not affect the motion of the polymer along the
tube. The second term is just like the random potential of the homopolymer
problem (\ref{homo}); it is irrelevant as discussed above.] The
Hamiltonian (\ref{dp1}) then describes a fictitious ``directed path'' with
``transverse'' coordinates $\left\{ u_n\right\} $, embedded in a $1+1$
dimensional random medium $W(u,n)$~\cite{note1}. The randomness results
from a combination of sequence heterogeneiity
and medium disorder. To elucidate the
properties of this system, we first perform a naive perturbative analysis
for small randomness $\Delta\equiv \Delta_q \Delta_V $. 
We find the randomness to have a negligible effect on
the classical reptation results for chains below a crossover
scale $N_{\times }\approx (k_BT)^5/(K\Delta ^2a^2)$. The perturbative
analysis fails for long chains with $N > N_{\times }$, indicating that 
the {\em asymptotic reptation properties are qualitatively affected by 
arbitrarily weak sequence heterogeneities}.
 Note the dependence of
the polymer's crossover length $M_{\times }\equiv N_{\times }\,M_{\rm blob}$ on
the tube size $a$: Using expressions for $K$ and $\Delta $ given above for
the simple RHP model, we find $M_{\times }\sim a^4b/\sigma^2\xi^5$, 
which increases quickly with increasing $a/\xi $ as suggested in 
Ref.~\cite{dg}, but is
nevertheless accessible for sufficiently heterogeneous polymers 
in typical gels.

To obtain the asymptotic behavior of the polymer with $M\gg M_\times$,
regime, let us examine more closely the effective random potential $W(u,n)$%
: The statistics of $W$ is easily obtained in terms of the statistics of $q_n
$ and $V_1(s)$, with $\overline{[W]}=0$ and 
\begin{equation}
\overline{\lbrack W(u,n)W(u^{\prime },n^{\prime })]}=\Delta \,\delta
_a(u-u^{\prime })\delta _{n,n^{\prime }}.  \label{WW}
\end{equation}
The  correlator (\ref{WW}) indicates that an effective {\em %
two-dimensional} random ``point'' potential is {\em generated}, even though $%
W$ itself, being a product of two {\em one-dimensional} random variables,
must contain long-range correlations. The latter is manifested in the higher
moments of $W$. For instance, $W^2(u,n)$ contains a term $\left[ (\delta
q)^2\right] \,\delta V_1^2(u_n+na)$ which is correlated along the direction $%
u_n=-na$. But such terms are just like the random potential $V(s)$ of the
homopolymer case and are irrelevant. Thus, we conjecture that {\em 
the system (\ref{dp1}) belongs to the universality
class of a directed path in (1+1)-dimensional uncorrelated Gaussian random
potential.}

This universality class is well known~\cite{dp}. In the asymptotic regime, 
the polymer exhibits glassy dynamics which is characterized
by two exact scaling laws: The sample-to-sample free energy variation scales
as $\delta F(N)\approx k_BT\cdot (N/N_{\times })^{1/3}$, 
and the contour length fluctuation scales as $\delta L(N)\approx s_{\times
}\cdot (N/N_{\times })^{2/3}$, where $s_{\times }=\delta L_0(N_{\times
})=(k_BT)^3/(K\Delta\, a)$. 
The conjectured equivalence between RHP reptation and the directed path
problem were tested, by numerically computing $\delta F(N)$
and $\delta L(N)$ from the model (\ref{homo}) and the interaction (\ref{U}),
treating $K$, $\Delta $ as arbitrary parameters. We used the
transfer matrix method~\cite{dp}, and examined particularly the
``zero-temperature'' limit of the model, whose small crossover length $%
N_{\times }$ enables us to access the asymptotic scaling regime. The
zero-temperature problem is an optimization problem. It is necessary to
place the bead-spring chain on a discrete lattice. To simplify the numerics,
we restrict the displacement of beads such that $s_{n+1}-s_n$ can only take
on the values $\{a,a\pm 1\}$ at each step $n$ of the transfer matrix. 
A random potential $V_1(s)$ of zero mean and unit variance is assigned to
each lattice point, and a charge $q_n$ is assigned independently to each
bead. The numerical results for the disorder-averaged contour length
fluctuation $\delta L=\overline{[(u_N^{*})^2]}^{1/2}$ of the optimal
configuration $u_n^{*}$ (with the $n=0$ end fixed at the origin) and the
sample-to-sample variation $\delta E^{*}(N)$ of the total energy of the
optimal configuration $E^{*}(N)$~\cite{E-min} are shown in Fig.~1(a). 
It is seen that
fluctuations rapidly approach those expected of the directed path,
$\delta L\sim N^{2/3}$ and $\delta E^{*}\sim N^{1/3}$, 
suggesting that the RHP is indeed in 
\begin{figure}
\epsfysize=1.5truein
\centerline{\epsffile{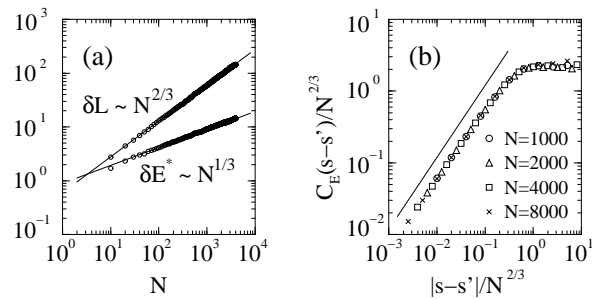}}
\vspace{10pt}
\caption{
Heteropolymers confined in a tube: 
(a) Scaling of the contour
length fluctuation $\delta L$, and the sample-to-sample minimal energy
variation $\delta E^{*}$, for bead-spring chains of $N=4000$, $a=4$, $K=1$,
and $[q]=0.1$, averaged over $1500$ samples.
(b) Auto-correlation of the energy landscape 
$E(s,N)$; the straight line has slope $1$.
}
\end{figure}
\noindent the same
universality class as the 1+1 dimensional directed path. 
Similar behaviors have recently been found in a number of related 
studies~\cite{hl,ch}.

Another useful quantity to examine is variation in the polymer's ``energy
landscape'', which we obtain by fixing the $n=0$ end of the polymer to an
arbitrary coordinate $s$, and then computing the energy $E(s,N)$ of the
optimal configuration (of the constrained polymer) for each $s$. The
landscape is characterized by the correlation function $C_E(s-s^{\prime
},N)\equiv \left[ \overline{(E(s,N)-E(s^{\prime },N))^2}\right] $, which is
expected~\cite{dp} to have the scaling form $C_E(s,N)=N^{2/3}g(|s|/N^{2/3})$%
, with the scaling function $g(x\lesssim 1)\sim x$ and $g(x\gtrsim 1)\sim 
{\rm const}$. This scaling form is verified by our numerics as shown in 
Fig.~1(b). 

The zero-temperature behavior described here corresponds to the asymptotic
regime of the RHP, beyond the crossover scale $N_{\times }$. 
For the finite temperature problem at hand, the RHP dynamics
along the tube is governed by
variations in the {\em free-energy} landscape $F(s,N)$, whose form is
obtained easily from $E(s,N)$. It is described by the correlation function 
$C_F(s-s^{\prime },N)=(k_BT)^2\left| s-s^{\prime }\right| /s_{\times }$, for
$s_{\times }<\left| s-s^{\prime }\right| <\delta L(N)$, saturating to 
$C_F\sim \left( \delta F\right) ^2\approx (k_BT)^2(N/N_{\times })^{2/3}$ for
larger displacements $\left| s-s^{\prime }\right| $. 

Dynamics of the chain in such a rough free energy landscape cannot proceed
by the propagation of a few isolated kinks. It requires 
instead {\em collective}
movement of large pieces of the chain such that the chain always remains in
local optimal configurations~\cite{mdk}. The resulting 
``collective creep" dynamics can be modeled 
by the motion of one of the chain ends 
in the 1d Brownian random potential $F(s,N)$~\cite{ldv}. The latter is a
well-studied classic problem~\cite{sinai}. The
time it takes for the particle to move by a distance $s>s_{\times }$ along
the tube is $\tau _{R,0}(N_{\times })\cdot \exp \left[
C_F(s)/(k_BT)^2\right] \sim e^{s/s_{\times }}$. Since $%
C_F(s)/(k_BT)^2$ saturates at $(N/N_{\times })^{2/3}$ for $\delta L<s<L$,
the reptation time is 
\begin{equation}
\tau _R(N)\approx \tau _{R,0}(N_{\times })\exp [(N/N_{\times })^{2/3}],
\label{tau_r}
\end{equation}
which greatly exceeds the classical reptation time $\tau _{R,0}(N)$ if the
polymer is sufficiently long. Note that $\tau _R$ also
exceeds the waiting time $\sim e^{N^\omega/k_B T}$
needed for the activated barrier hopping by the
homopolymer. Hence we conclude that {\em the dynamics of a long random
heteropolymer is much slower than that of a homopolymer}. 

The extraordinary slow dynamics of the RHP is
a {\em collective phenomenon} resulting from a global ``resonance'' of the
randomness in the polymer composition with the randomness in the tube
structure; it is not a local effect, say, an increase of the microscopic
friction. This can be tested by comparing an RHP
with a $ABAB$ copolymer consisted of the same monomers.
Upon coarse-graining by a
finite scale, the  copolymer becomes equivalent to a homopolymer,
while heterogeneities of an RHP cannot be coarse-grained away. Thus we
expect a long RHP 
to exhibit much slower dynamics than a periodic  copolymer
of the same length and monomer content. 

We are grateful to David R. Nelson and Bruno H.~Zimm for helpful
discussions. We are particularly indebted to Prof.~Nelson for an English
translation of Ref.~\cite{dg}. This work is supported by an ONR Young
Investigator Award, and the Petroleum Research Fund through grant 
ACS-PRF \# 32517-G7.
TH also acknowledges the support of an A.P. Sloan
research fellowship.



\begin{references}
\bibitem{note1} Note that de Gennes's restricted model~\cite{dg} corresponds to
a directed path in $(0+1)$~dimension.
\bibitem{E-min}  For convenience, we monitor the energy difference between $%
E^{*}(N)$ and the energy of the optimal configuration with the other ($n=N$)
end also fixed at $u_N=0$.
\end{references}
\end{document}